\documentclass[reprint,superscriptaddress,amsmath,amssymb,aps,pra]{revtex4-2}

\usepackage{graphicx}
\usepackage{dcolumn}
\usepackage{bm}
\usepackage{epstopdf}
\usepackage{braket}
\usepackage{color}
\usepackage{mathrsfs}
\usepackage{ragged2e}
\usepackage{eqnarray}
\usepackage{subeqnarray}
\usepackage{cases}
\usepackage{amsfonts}
\usepackage{dcolumn}
\usepackage{bm}
\usepackage{tikz}
\usepackage{float}
\usepackage[colorlinks=true,citecolor=blue,urlcolor=blue]{hyperref}
\usepackage{amsmath,amsfonts,amssymb,times,natbib}
\usepackage[standard]{ntheorem}
\usepackage{natbib}
\usepackage{stmaryrd}
\usepackage{comment}

\begin{document}

\preprint{APS/123-QED}

\title{Transient Dynamics and Duty-Cycle Optimization in Pulse-Width-Modulated Bell–Bloom Pumping}

\author{Ying-Hao Ye}
\email{yyhustc@mail.ustc.edu.cn}
\affiliation{School of Physics, Hefei University of Technology, Hefei, Anhui 230009, China.}%

\author{Ling-Yan Hu}
\affiliation{School of Physics, Hefei University of Technology, Hefei, Anhui 230009, China.}%

\author{Dui-Gao Yi}
\affiliation{School of Physics, Hefei University of Technology, Hefei, Anhui 230009, China.}%

\author{Zhi-Fei Yu}
\email{zfyuphy@hfut.edu.cn}
\affiliation{School of Physics, Hefei University of Technology, Hefei, Anhui 230009, China.}%
\affiliation{Key Laboratory of Opto-Electronic Information Acquisition and Manipulation (Anhui University), Ministry of Education, Hefei, Anhui 230039, China.}%

\author{Bing Chen}
\email{bingchenphysics@hfut.edu.cn}
\affiliation{School of Physics, Hefei University of Technology, Hefei, Anhui 230009, China.}


\date{\today}

\begin{abstract}
We investigate the duty-cycle-dependent atomic response under full-depth intensity pulse-width modulation in Bell--Bloom optical pumping. A time-domain Bloch model explicitly resolves the pump-on/off spin dynamics, yielding piecewise analytical transient solutions and a periodic steady-state description without cycle averaging or harmonic truncation. An extended free-induction-decay method independently determines the effective dark and pump-on transverse-relaxation rates, $W_0$ and $W$. The predicted lock-in responses agree well with experimental results. We find that for each Larmor frequency $\omega_0$, the slope is maximized at a finite pump rate that increases with $\omega_0$. These results provide a practical framework for sensitivity-oriented optimization of the duty cycle and pump rate in PWM-driven atomic magnetometers.
\end{abstract}

\maketitle


\section{\label{sec:introduction}Introduction}

Finite-field optically pumped magnetometers infer the magnetic-field magnitude from the Larmor precession of ground-state atomic polarization \cite{np07OPMreview,Fabricant2023}. In resonantly driven implementations, spin coherence is sustained either by an oscillating magnetic field, as in $M_x/M_z$ magnetometers, or by synchronous optical pumping, as in the Bell--Bloom magnetometer (BBM) \cite{BellBloom61,Bloom62}. The BB scheme is magnetically silent and avoids coil crosstalk, misalignment errors, and rf-induced power broadening. Under comparable optimized conditions, it can retain a sensitivity similar to that of an $M_x$ magnetometer while exhibiting fewer directional dead zones, making it promising for compact and multichannel sensing \cite{SensitivityComparisonofMx}. To date, fT- and sub-fT-level sensitivities in optically pumped magnetometers have been achieved by prolonging spin coherence with antirelaxation coatings \cite{ParaffinBalabas06,DingDualSpeciesPRApplied2023,RosnerAPL2022} or buffer gases \cite{FemtoteslaOE2010,MhaskarlowpowermagnetometerAPL2012,RushtonAlignmentBasedPRApplied2023} and by suppressing spin-exchange relaxation in the low-field SERF regime \cite{SERF1,SERFexp,nature03subft}. 

In the small-rotation regime, the balanced-polarimeter signal in a BBM is proportional to the projection of the atomic polarization along the probe propagation direction \cite{magnetoopticalreview}, directly connecting the measured optical response to the Bloch description of spin precession and relaxation \cite{nuclearinduction}. Building on this framework, Bell and Bloom demonstrated that spin precession can be resonantly driven through synchronous optical pumping, a mechanism that establishes the foundation of BBMs \cite{BellBloom61,Bloom62}. The synchronous pumping can be implemented by modulating the light intensity \cite{AMGawlik06}, frequency \cite{FMBudker}, or polarization \cite{PMOE2013}. These schemes can be described within a common Fourier framework in which the resonance strengths are determined by the spectral components of the light-matter interaction \cite{GrujicWeis2013}. Among these schemes, intensity modulation (IM) is attractive for its direct control of the instantaneous pumping rate; in a two‑beam configuration with pump and probe separated, IM suppresses direct feedthrough and delivers a clean signal well suited for self‑oscillating feedback \cite{Higbie06}. At a modulation depth of $100\%$, pulse-width modulation (PWM) produces a binary on-off pulse train that can be generated without high-resolution waveform synthesis, thereby potentially simplifying the control electronics and facilitating sensor miniaturization. These considerations motivate the present focus on full-depth intensity PWM and the dependence of the BB-type response on its duty cycle.

IM magneto-optical resonances and their multiharmonic structure have been investigated experimentally and theoretically \cite{AMGawlik06,GrujicWeis2013}. Duty-cycle effects have also been studied for several modulation schemes \cite{PMOE2013}. In separated pump-probe IM magnetometers operated at fixed average pump power, shorter pump pulses were experimentally found to improve performance under the conditions studied, with the best result occurring at the shortest tested duty cycle \cite{Gerginov2017}. A more insightful result was obtained by Wang \textit{et al.} for IM with fixed peak pump rate \cite{Wang2016}. They showed that, within a treatment where the pump-induced relaxation is represented by its cycle-averaged value, the maximum field response occurs when $\eta R_0 = 1/T_2$, where $R_0$ is the pump-on optical-pumping rate and $T_2$ is the transverse-relaxation time without pump light. This cycle-averaged treatment, however, does not explicitly resolve the distinct spin dynamics during the pump-on/off stages. This distinction may become important when the instantaneous pumping rate approaches or exceeds the characteristic rates of Larmor precession and relaxation. At such rates, time-dependent pump-induced relaxation may couple Fourier components and alter the phase accumulated over a cycle \cite{GrujicWeis2013,Floquet}, making the average approximation insufficient to characterize the response. Although a frequency-domain treatment remains possible, numerical convergence may require a large harmonic basis, especially for short, intense pulses, thereby increasing required computation resource. 

In this work, we develop a time-domain model that explicitly resolves the pump-on/off spin dynamics under full-depth PWM. 
This treatment retains the full time dependence of pump-induced relaxation without cycle averaging or harmonic-basis truncation. It therefore extends the description to regimes in which the intraperiod dynamics produced by strong pulsed pumping cannot be neglected. To connect theory and experiment, we introduce an extended free-induction-decay (eFID) framework, building on conventional FID \cite{FID2018}, that determines the pump-on/off transverse-relaxation rates independently of the steady-state response. 
As shown in Ref.~\cite{Wang2016}, the duty cycles optimizing the resonance amplitude and zero-crossing slope generally differ; we further show that the pumping rate optimizing the slope depends on the Larmor frequency. The model recovers this cycle-averaged scaling in the short-duty-cycle limit and quantifies deviations as the pump-on evolution becomes appreciable. Together, the time-domain model and the proposed parameter-calibration protocol constitute an integrated theoretical and experimental toolkit for optimization of PWM BB type operation.

\section{\label{sec:responseandreadout}Atomic Response and Readout}
\begin{figure}[th!]
	\centering
	\includegraphics[width=\linewidth]{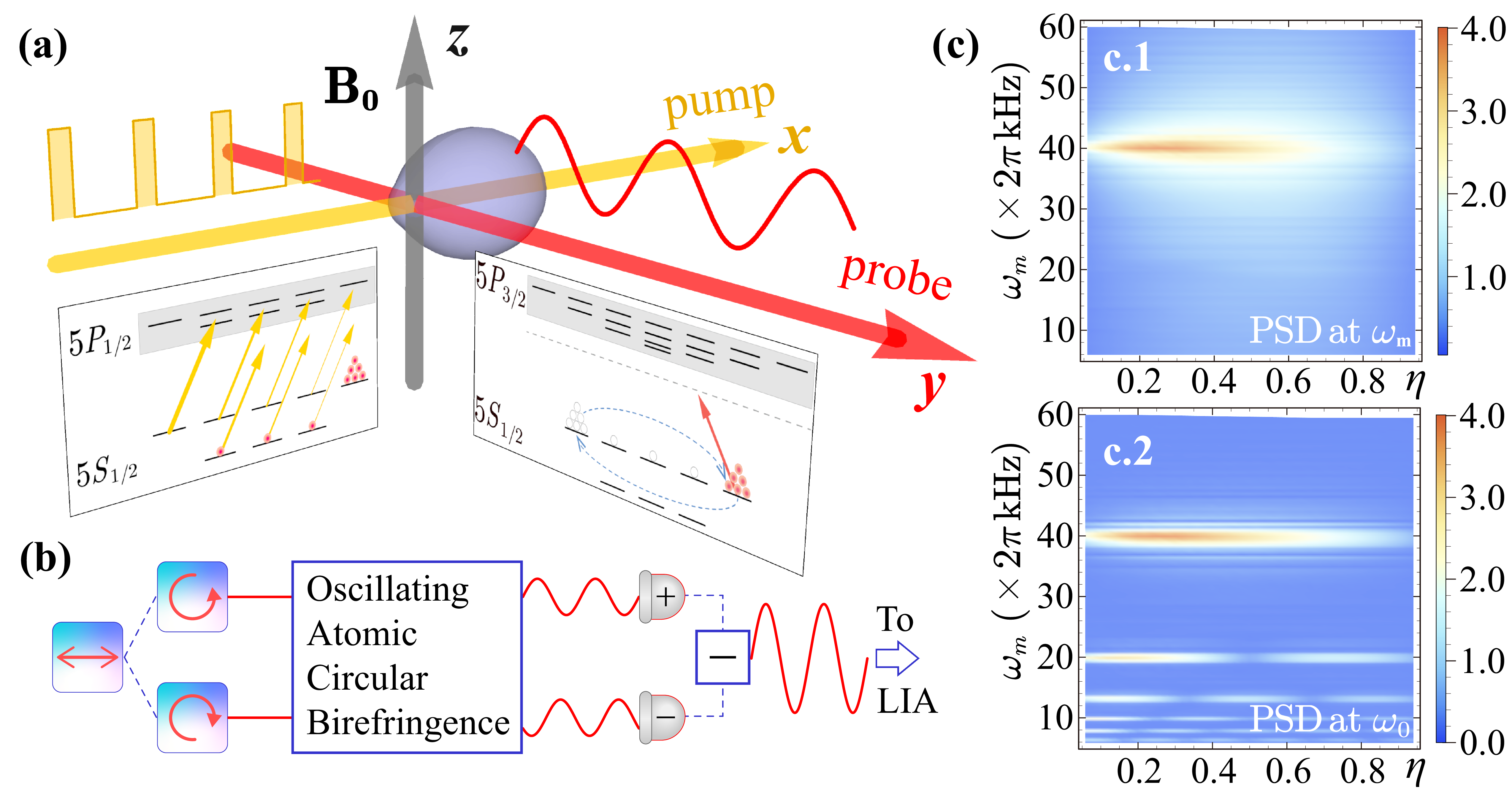}
	\caption{Principle of the PWM Bell–Bloom excitation. (a) Schematic of the experimental geometry, together with the corresponding pump and probe transition schemes. (b) Polarimeter detection principle based on magneto-optical rotation. (c) c.1 Power spectral density (PSD) evaluated at the pump modulation frequency as a function of the modulation frequency $\omega_m$ and pump duty cycle $\eta$. c.2 PSD evaluated at the Larmor frequency $\omega_0$ under the same conditions.
	}
	\label{Fig:principle}
\end{figure}
Figure~\ref{Fig:principle} (a) and (b) illustrates the overall scheme of Bell--Bloom excitation using full-depth intensity PWM and the differential polarimeter readout. A circularly polarized pump beam addressing the $\mathrm{D_1}$ transition of $\mathrm{^{87}Rb}$ optically pumps the atomic ensemble. Owing to the selection rules, the atomic population is gradually transferred to a stretched Zeeman sublevel, resulting in the collective spin polarization, which can be visualized by an example angular-momentum probability surface (AMPS) \cite{AMPS} shown in Fig.~\ref{Fig:principle}(a). 
The spin polarization component along the probe direction $P_y$ induces circular birefringence for probe light, and the resulting magneto-optical rotation is converted by the balanced photodetector into a differential signal whose time-dependent component is proportional to $P_y$, as illustrated in Fig.~\ref{Fig:principle}(b):
\begin{equation}
	\label{eqn:VdproptoPy}
	V_d(t)\propto P_y(t).
\end{equation}

A pump beam periodically drives the atomic medium at an angular frequency $\omega_m$, while the collective magnetic moment associated with spin polarization undergoes Larmor precession about the external field $\mathbf{B}_0$ at $\omega_0=\gamma B_0$ during the pump-off intervals. Their interplay yields a polarimeter signal with frequency $\omega_m$. The power spectral density (PSD) distribution of the detected signal, shown in Fig.~\ref{Fig:principle}(c), gives an intuitive overview of the magneto-optical response. The PSD evaluated at $\omega_m$ exhibits a pronounced maximum around $\omega_m=\omega_0$, with a strong dependence on pump duty cycle $\eta$ and a clear optimum. In contrast, the PSD evaluated at $\omega_0$ not only exhibits a dominant resonance at $\omega_m=\omega_0$, but also reveals clear additional maxima at $\omega_m=\omega_0 /n , \, n=2,3,\ldots$, directly corroborating that the resolved Fourier harmonics of the pump waveform also participate in the spin dynamics \cite{AMGawlik06,GrujicWeis2013}. Along each $n$th-order resonance, the duty-cycle ($\eta$)-dependent minima align with $\eta=k/n,\,k=1,\ldots,n-1$, where the corresponding square-wave Fourier coefficient vanishes. These observations reveal a practical limitation: frequency-domain treatments may require retaining many harmonics, especially under strong pumping. In time domain, these resonances arise from stroboscopic phase matching: at $\omega_m=\omega_0/n$, successive pump cycles coherently reinforce the spin polarization every $n$ Larmor periods. This picture motivates the periodic-boundary formulation developed below.

\section{\label{sec:Blochmodel}Transient Dynamics and Extended FID Framework}
\begin{figure*}[ht!]
	\centering
	\includegraphics[width=\linewidth]{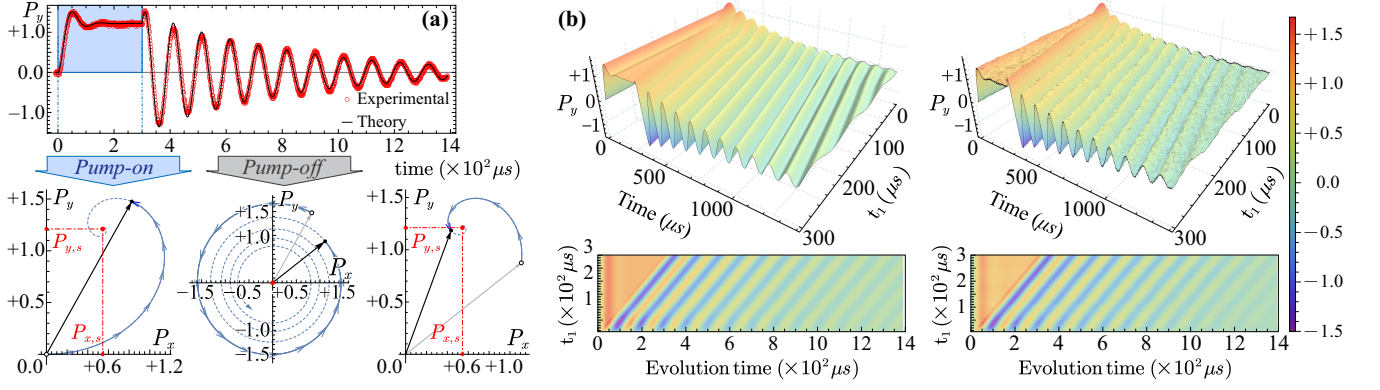}
	\caption{Visualization of atomic spin-polarization precession. (a) Extended FID-based characterization of the pump-on and pump-off spin dynamics. The lower panels show the corresponding spin-polarization vector trajectory in the $P_x$-$P_y$ plane. 
		The fit yields $T_2=\text{526}\,\mu s$, $W_0=\text{1.9}\,\text{kHz}$, and $W=\text{27.4}\,\text{kHz}$ at $\text{130}\,^\circ\mathrm{C}$. (b) Robustness test of the eFID characterization for different $t_1$. The left panels show the calculated $P_y$ using the averaged fitted parameters, and the right panels show the measured transient response. The same colormap is used for the calculated and measured data. The averaged fitting results over the full $t_1$ scan are $W_0=1.898\pm0.005 \, \mathrm{kHz}$ and $W=27.9\pm0.5 \,\mathrm{kHz}$.}
	\label{Fig:atomicprecession}
\end{figure*}
Given the Doppler- and pressure-broadened optical response at temperatures above $130\,^\circ\mathrm{C}$ and with $760$ Torr of $N_2$ buffer gas, we parameterize the net pumping dynamics by the effective optical-pumping and pump-induced-relaxation rates, $R_{\mathrm{op}}$ and $\xi R_{\mathrm{op}}$, in the following Bloch equation \cite{effectivemasterequations}:
\begin{equation}
	\label{eqn:Blocheq}
	\dot{\mathbf{P}}=\omega_{0}\hat{\mathbf{e}}_{B}\times \mathbf{P}+R_{\rm op} \hat{\mathbf{e}}_{\rm p}-\xi R_{\rm op} \mathbf{P}-\frac{\mathbf{P}_{\perp}}{T_2}-\frac{\mathbf{P}_{\sslash}}{T_1}.
\end{equation}
Here, $T_1$ and $T_2$ denote the longitudinal and transverse relaxation times, respectively. $\hat{\mathbf{e}}_{B}$ and $\hat{\mathbf{e}}_{\rm p}$ are the unit vectors along the magnetic field and the pump direction. In a typical two-beam BB configuration, the magnetic field, pump, and probe directions are mutually orthogonal. Without loss of generality, we take $\hat{\mathbf{e}}_{B}=\hat{\mathbf{e}}_{z}$ and $\hat{\mathbf{e}}_{\rm p}=\hat{\mathbf{e}}_{x}$, Eq.~\ref{eqn:Blocheq} is then expanded in Cartesian coordinates as:
\begin{subequations}
	\label{eqn:dPdtoriginal}
	\begin{eqnarray}
		\dot{P}_x&=&-P_x/T_2-\xi P_x R_{op}-\omega_{0} P_y+R_{op}, \label{dPxdt}
		\\
		\dot{P}_y&=&-P_y/T_2-\xi P_y R_{op}+\omega_{0} P_x. \label{dPydt}
	\end{eqnarray}
\end{subequations}
Since the probe measures $P_y$ (Eq.~\ref{eqn:VdproptoPy}), only the coupled transverse dynamics of $P_x$ and $P_y$ need to be retained, therefore leaving $T_2$ as the only intrinsic relaxation time. The general transient solution to Eqs.~\ref{eqn:dPdtoriginal} with pumping reads: 
\begin{subequations}
	\label{eqn:transientsolutionconstantpump}
	\begin{align}
		P_x(t)
		&= P_{x,s}-e^{-Wt}
		[\Delta P_x\cos\omega_0t-\Delta P_y\sin\omega_0t],
		\label{Pxtransientconstantpump}\\
		P_y(t)
		&= P_{y,s}-e^{-Wt}
		[\Delta P_x\sin\omega_0t+\Delta P_y\cos\omega_0t].
		\label{Pytransientconstantpump}
	\end{align}
\end{subequations}
Here, $W=1/T_2+\xi R_{\rm op}$ stands for the effective relaxation rates during pump-on stage and $\Delta P_{i}=P_{i,s}-P_{i,0}|_{i=x,y}$ denotes the deviation of the initial spin polarization $P_{i,0}$ from its corresponding steady-state value $P_{i,s}$. The above equations describe the optically driven spin evolution, with $W$ governing the relaxation and $\omega_0$ the Larmor precession, toward the corresponding steady state.
\begin{equation}
	\label{eqn:PxsPys}
	P_{x,s}=\frac{W R_{\rm op}}{W^2+\omega_0^2};\quad
	P_{y,s}=\frac{\omega_0 R_{\rm op}}{W^2+\omega_0^2}
\end{equation}
Upon evolving under a constant pump for a duration $t_1$, the pumping light is switched off, i.e. $R_{\rm op}=0$, and the transverse polarization undergoes free Larmor precession with relaxation rate $W_0=1/T_2$:
\begin{subequations}
	\label{eqn:transientsolutionnopump}
	\begin{align}
		P_x^{\prime} \left( \delta t ,t_1\right) &= e^{-W_0 \delta t} \left[P_{x}\left(t_1\right)\mathrm{cos}\omega_0 \delta t-P_y\left( t_1 \right) \mathrm{sin}\omega_0 \delta t\right], \label{Pxfreeevo}\\
		P_y^{\prime} \left( \delta t ,t_1\right) &= e^{-W_0 \delta t} \left[P_{x}\left(t_1\right)\mathrm{sin}\omega_0 \delta t+P_y\left( t_1 \right) \mathrm{cos}\omega_0 \delta t\right]. \label{Pyfreeevo}
	\end{align}
\end{subequations}
Here $\delta t=t-t_1$ is the free precession duration. The analytical solutions Eqs.~\ref{eqn:transientsolutionconstantpump} and \ref{eqn:transientsolutionnopump} provide a quantitative framework for characterizing the relaxation rate at pump-on ($W$) and pump-off ($W_0$) stages simultaneously. Under weak-probe conditions, where the probe is blue-detuned by approximately $30~\mathrm{GHz}$ from $\mathrm{^{87}Rb}$ $\mathrm{D_2}$ transition and limited to a peak power below $1~\mu\mathrm{W}$, we extend the conventional FID-based characterization by incorporating both the pump-on and pump-off stages into a single self-consistent framework (eFID). Pump pulses of duration $t_1$ are repeated with a period much longer than $T_2$, allowing residual transverse polarization to decay between cycles, while an oscilloscope records the polarimeter transient (see Fig.~\ref{Fig:atomicprecession}(a) for example). First, $W_0$ and $\omega_0$ are extracted from the pump-off segment using Eq.~\ref{Pyfreeevo}. These values are then used with Eq.~\ref{Pytransientconstantpump} to determine $W$ from the pump-on segment. Finally, a joint fit to both stages enforces continuity at the switching time and yields self-consistent values of $W_0$ and $W$. As shown in Fig.~\ref{Fig:atomicprecession}(a), the recorded polarimeter signal is well reproduced by the fitted response over the entire sequence and the lower panels visualize the fitted spin polarization in the $P_x$-$P_y$ plane.
\begin{figure}[htb!]
	\centering
	\includegraphics[width=180pt]{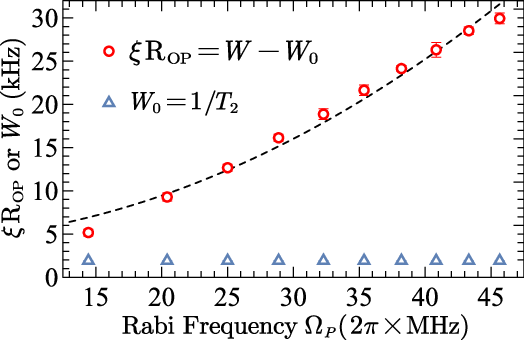}
	\caption{Pump-induced relaxation $W-W_0$ and intrinsic relaxation rate $W_0$ extracted using the eFID framework as functions of the pump Rabi frequency. The dashed curve shows the result obtained from a quadratic fit (assuming $\xi R_{op}\propto \Omega_p^2$).}
	\label{Fig:pumpRabifrequncy}
\end{figure}

The robustness of the extracted parameters was examined by repeating the extended eFID fitting for different $t_1$, ranging from 5 $\mu s$ to 300 $\mu s$. This range covers pump durations much shorter than and longer than one Larmor period. The relative standard deviations for all parameters were below 5$\%$ over the scan range, indicating that the extracted parameters are robust against the choice of $t_1$. As shown in Fig.~\ref{Fig:atomicprecession}(b), The calculated and measured two-dimensional transient responses agree well. In the subsequent measurements, the same eFID characterization was performed before and after each experimental run to determine the dynamic parameters and to monitor the system stability.

By utilizing the proposed framwork, we measured $W$ and $W_0$ as functions of the pump Rabi frequency $\Omega_p$. As shown in Fig.~\ref{Fig:pumpRabifrequncy}, the pump-induced relaxation $\xi R_{\rm op}=W-W_0$ scales approximately quadratically with $\Omega_p$ over the investigated range, whereas $W_0=1/T_2$ remains nearly unchanged. These results indicate that the spin relaxation in the pump-off stage barely perpetuated by the pump light and justify the use of $W$ and $W_0$ as independent effective parameters in the following analysis.

\section{\label{sec:periodicsolution}Periodic Solution and Theoretical analysis}
\begin{figure}[ht!]
	\centering
	\includegraphics[width=\linewidth]{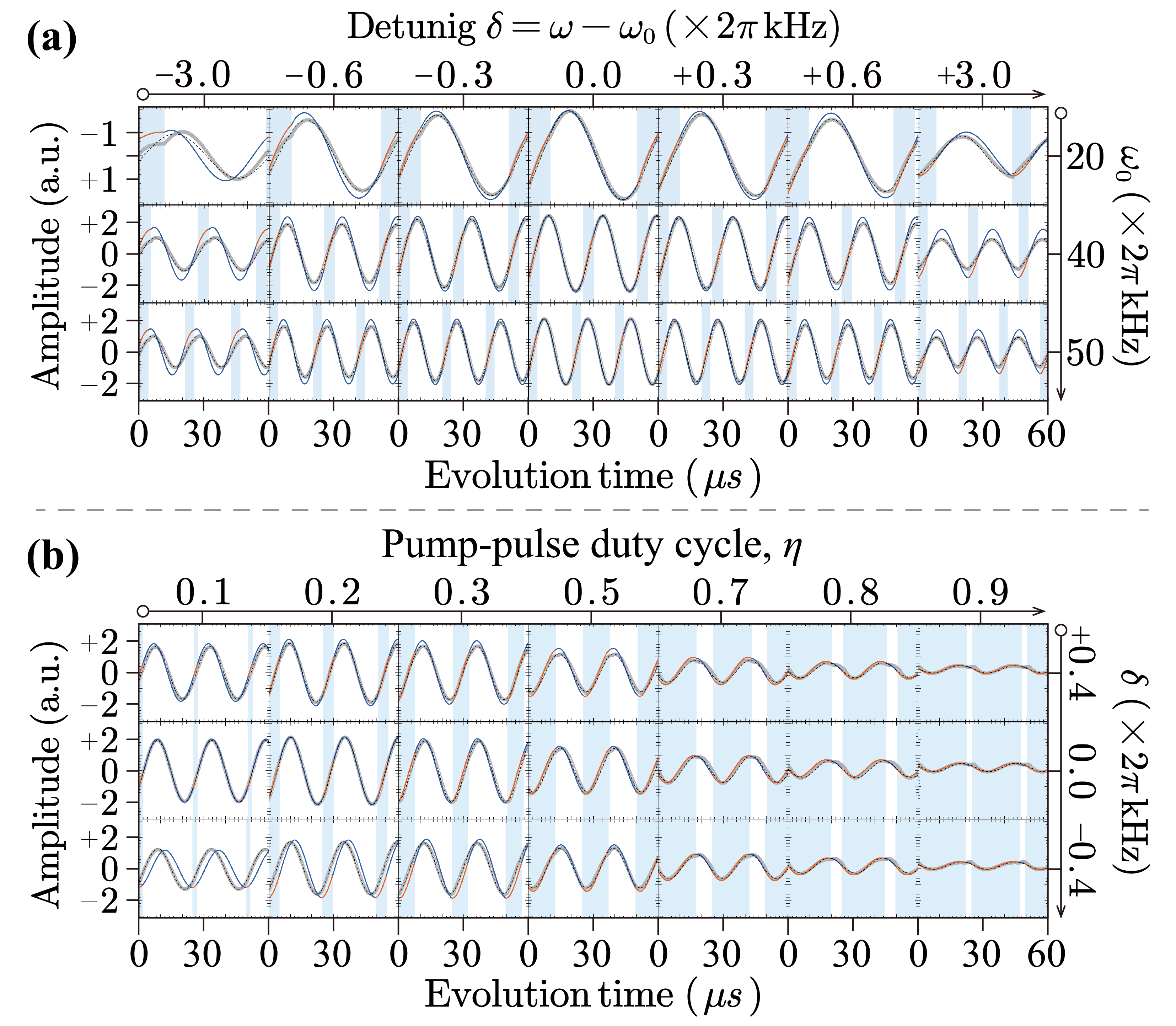}
	\caption{Representative polarimeter signals measured under different modulation conditions. Gray dots denote the oscilloscope traces, orange and blue solid curves show the calculated waveforms, and black dashed curves represent sinusoidal fits. The light-blue shaded regions indicate the pump-on duration.}
	\label{Fig:waveform}
\end{figure}
Under periodic pumping, the system reaches a periodic steady state in which the spin polarization at the beginning of each period is reproduced after one complete pump-on/off cycle. This boundary condition gives
\begin{equation}
	\label{eqn:periodicboundarycondition}
	\mathbf{P}_{\perp}(t=0)=\mathbf{P}_{\perp}^{\prime}(T-\eta T,\eta T).
\end{equation}
Here, $\mathbf{P}_{\perp}(t=0)=P_{x,0}\hat{\mathbf{e}}_x+P_{y,0}\hat{\mathbf{e}}_y$ is the initial transverse spin polarization of each cycle, and $\mathbf{P}_{\perp}^{\prime}(T-\eta T,\eta T)$ is its value at the end of one period, evaluated from Eqs.~\ref{eqn:transientsolutionnopump} with $T=2\pi/\omega_m$. The periodic boundary condition forms a set of linear equations for $P_{x,0}$ and $P_{y,0}$, thereby determining the stable periodic spin trajectory, which can be fully predicted from the calibrated parameters $W$ and $W_0$ through Eqs.~\ref{eqn:transientsolutionconstantpump} and \ref{eqn:transientsolutionnopump}.
Comparisons between theoretically predicted and measured waveforms for several representative cases are summarized in FIG.~\ref{Fig:waveform}. For these measurements, the sample temperature was raised to $162\,^\circ\mathrm{C}$ to prolong the transverse relaxation time. The extended FID fitting framework yields $W=32\pm2\,\mathrm{kHz}$ and $W_0=0.6\pm0.2\,\mathrm{kHz}$ at $\omega_0=2\pi\times40\,\mathrm{kHz}$.

Fig.~\ref{Fig:waveform}(a) compares the measured and calculated waveforms at $\eta=0.2$ for several bias fields and detunings, $\delta\omega_m=\omega_m-\omega_0$. Near resonance, the measured traces are well reproduced by the calculated waveforms and are nearly sinusoidal. This agreement over Larmor frequencies from $20$ to $50\,\mathrm{kHz}$ demonstrates that the present model remains valid over a wide range of bias magnetic fields. As the detuning increases ($|\delta\omega_m|>W_0$), however, the measured waveforms gradually deviate from the predictions. Fig.~\ref{Fig:waveform}(b) shows that the waveform amplitude varies strongly and nonmonotonically with duty cycle. The near-resonant traces remain well described by both the theoretical calculations and sinusoidal fits, whereas larger $|\delta\omega_m|$ lead to increasing deviations from the predictions. 

The increasing deviation at larger detunings can be attributed to nonideal effects neglected in the treatment, such as pump-induced light shifts, nonuniform spatial profiles of the pumping/probing and relaxation rates, finite switching edges. As $|\delta \omega_m|$ increases, the Larmor phase accumulated over one modulation period deviates increasingly from $2\pi$, making the steady-state waveform more sensitive to additional phase shifts and amplitude distortions produced by these nonideal mechanisms. Nevertheless, typical BB-type applications are operated in the near-resonant regime, where the waveform is well described by the present model. Therefore, the observed off-resonant deviations have limited influence on the following analysis.
\begin{figure}[th!]
	\centering
	\includegraphics[width=\linewidth]{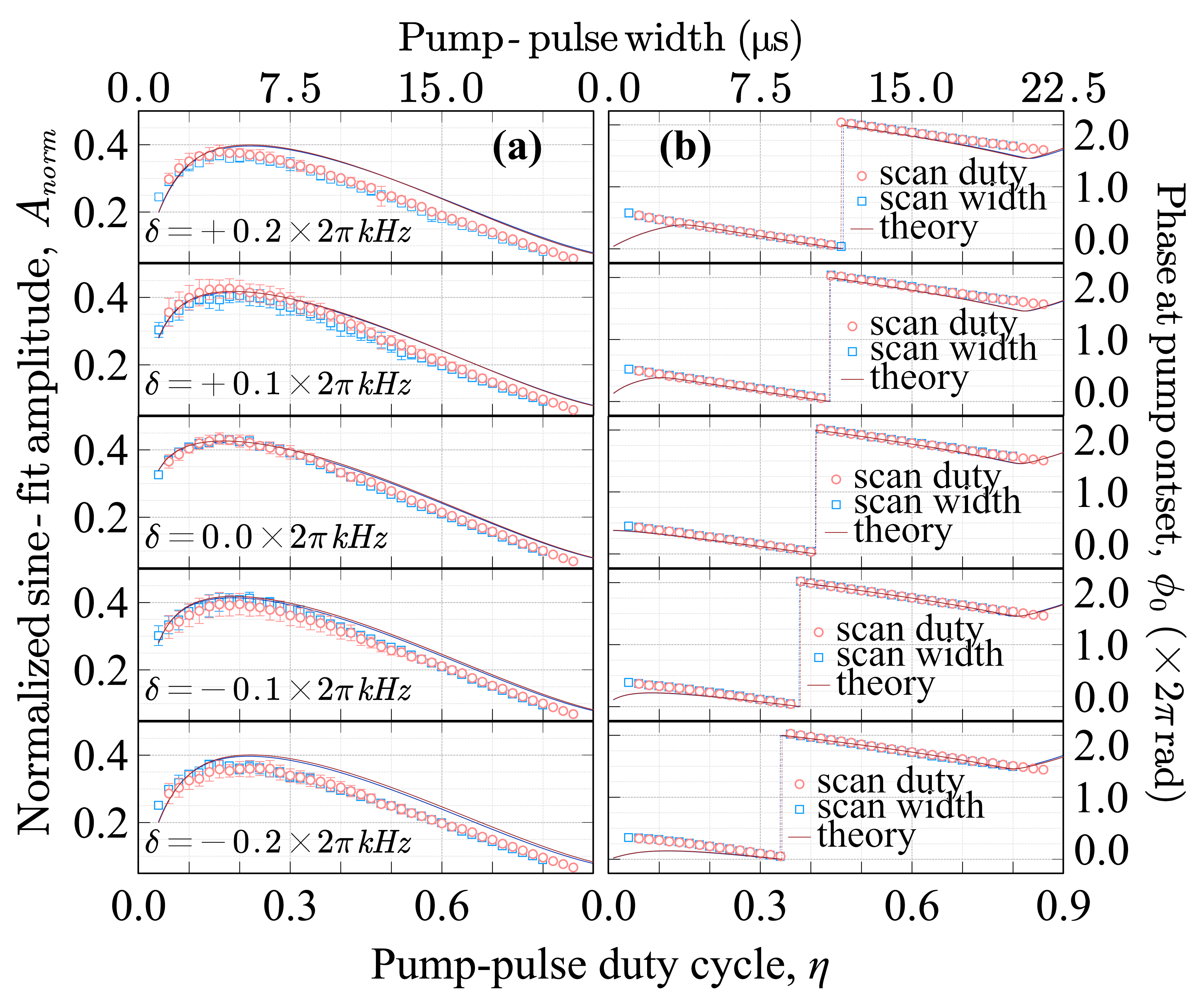}
	\caption{Sine-fit amplitude and phase extracted from the measured waveforms. (a) Normalized sine-fit amplitude $A_{\rm norm}$, obtained by normalizing the fitted amplitude to the pump-blocked probe sum signal, as a function of duty cycle $\eta$ and detuning $\delta\omega_m$. (b) Fitted phase $\phi_0$ compared with the theoretical prediction of Eq.~\ref{eqn:phi0}.
	}
	\label{Fig:amplitudeandphasefit}
\end{figure}

Motivated by the waveform comparison in FIG.~\ref{Fig:waveform} and the pioneering work \cite{Wang2016}, we focus on the range $\eta\le 0.5$, where the modulation response reaches its maximum. Under the small transverse relaxation condition $2\pi W_0\ll \omega_0$, the pump-off solution Eqs.~\ref{eqn:transientsolutionnopump} can then be approximated as $P_y^{\prime}(\tau,\eta T)
\simeq
P_\perp(\eta T)\sin(\omega_0\tau+\phi_1)$, where $\tau=t-\eta T$ and
\begin{equation}
	\label{eqn:Pvert}
	P_\perp(\eta T)^2=P_x^2(\eta T)+P_y^2(\eta T)
\end{equation}
corresponds to the transverse spin polarization prepared at the end of each pump pulse. Thus, according to Eq.~\ref{eqn:VdproptoPy}, the sine-fit amplitude is expected to scale with $P_\perp(\eta T)$, while the fitted phase $\phi_0$ at $t=0$ is determined by the orientation of the transverse spin vector at the pump-off/on switching point:
\begin{equation}
	\label{eqn:phi0}
	\phi_0 = \arcsin \left[\frac{P_{y,0}}{P_{\perp}(t_1)}\right].
\end{equation}
Solving Eqs.~\ref{eqn:periodicboundarycondition} and substituting Eqs.~\ref{eqn:transientsolutionconstantpump}-\ref{eqn:PxsPys} into Eq.~\ref{eqn:Pvert}, after rearrangement, we obtain the normalized amplitude of oscillation:
\begin{flalign}
	\label{eqn:Pverttoamplitude}
	&A_{\rm theo}^2 ( W_0,W,\omega_0,\eta) = P_{\perp}^2(t_1)/ P_{y,s}^2
	\nonumber
	\\
	&= \frac{W^2+\omega_0^2}{\omega_0^2}\cdot\frac{1+e^{-\frac{4\pi W\eta}{\omega_0}}-2 e ^{-\frac{2\pi W\eta}{\omega_0}}\cos (2\pi \eta)}{\left[1-e^{- \frac{2\pi W\eta}{\omega_0}- \frac{2\pi W_0}{\omega_0}(1-\eta)} \right]^2}.
\end{flalign}

To assess the ranges of validity of the predicted amplitude and phase, we summarize the experimental–theoretical comparison in Fig.~\ref{Fig:amplitudeandphasefit}. The measured amplitudes are normalized by the pump-off signal, yielding the dimensionless normalized sine-fit $A_{\rm norm}$. For each $(\eta,\delta\omega_m)$ pair, $A_{\rm norm}$ is compared with Eq.~\ref{eqn:Pverttoamplitude}, up to a global scale factor, as shown in Fig.~\ref{Fig:amplitudeandphasefit}(a). Similarly, the fitted phase was compared with the theoretical initial phase $\phi_0$ given by Eq.~\ref{eqn:phi0}, with an additional global phase offset accounting for electronic and timing delays, as shown in Fig.~\ref{Fig:amplitudeandphasefit}(b). For $\eta>0.5$, the pump-induced relaxation rate ($W \sim \omega_0$ under the present conditions) influences the resulted response strongly, leads to a more pronounced measurement--theory discrepancies. Within the magnetometer operating regime $\lvert\delta\omega_m\rvert\ll\omega_0$, the phase varies approximately linearly with $\eta$ over the duty-cycle range considered.

The quadrature output of a lock-in analysis is determined jointly by the signal amplitude and its relative phase with respect to the reference. By choosing the reference phase as the phase at resonance, the normalized demodulated output can be written as:
\begin{equation}
	\label{eqn:virtualLIAOutput}
	T_{\rm LIA}(\omega)=A_{\rm norm}(\omega)\sin\left[\phi_0 (\omega)-\phi_0(\omega_0)\right].
\end{equation}
The above equation vanishes at resonance and converts small detunings around $\omega_0$ into a dispersive-like response.

In a practical BB-type operation, measurements are obtained from the lock-in signal $T_{\rm LIA}$ at different magnetic fields $\gamma\omega_0$ under a fixed modulation frequency $\omega_m$. Equivalently, the magnetic-field response can be evaluated by fixing the bias magnetic field and sweeping $\omega_0$:
\begin{equation}
	\label{eqn:deltaBtodeltaomega}
	\left\lvert \frac{\partial T_{\rm{LIA}}}{\partial B_0}\right\rvert _{\delta =0}= \left\lvert\gamma\frac{\partial T_{\rm{LIA}}}{\partial \delta}\right\rvert _{\delta =0}=\gamma S_{\rm{LIA}}.
\end{equation}
Consequently, the near-resonance slope of $T_{\rm sim}$ with respect to the detuning characterizes frequency-to-output conversion gain of the magnetic-field response and thus $S_{\rm LIA}$ is one of the key figures of merit evaluating response sensitivity 

In this work, $S_{\rm{LIA}}$ in is estimated using a central difference method with a finite detuning interval $\Delta\omega=2\pi\times0.1\,\mathrm{kHz}$. To obtain an appropriate theoretical metric for duty-cycle optimization, we seek the theoretical differential limit of Eq.~\ref{eqn:deltaBtodeltaomega}. By substituting the exact periodic steady-state solutions of Eqs.~\ref{eqn:periodicboundarycondition} and \ref{eqn:phi0} into Eq.~\ref{eqn:deltaBtodeltaomega} we obtain a expression that is too cumbersome to provide physical insight. Under the experimentally relevant condition $W_0\ll\omega_0$, we neglect higher-order corrections associated with weak pump-off relaxation while retaining the leading decay factors that determine the periodic steady-state response. The slope then reduces to:
\begin{flalign}
	\label{eqn:Slopeideal}
	S_{\rm{LIA}}
	&\equiv
	\left|
	\frac{\partial T_{\mathrm{LIA}}}{\partial \delta}
	\right|_{\delta=0}
	&& \nonumber
	\\
	&=
	\frac{1}{\sqrt{1-P_{y,0}^{2}/P_{\perp}^{2}}}
	\left|
	\left[
	\frac{\partial P_{y,0}}{\partial \delta}
	-
	\frac{1}{2}
	\frac{P_{y,0}}{P_{\perp}^{2}}
	\frac{\partial P_{\perp}^{2}}{\partial\delta}
	\right]
	\right|_{\delta=0}
	&& \nonumber
	\\
	&\simeq
	\frac{2\pi}{\omega_0}
	\frac{
		\sqrt{
			1+e^{-\frac{4\pi W\eta}{\omega_0}}
			-2e^{-\frac{2\pi W\eta}{\omega_0}}
			\cos(2\pi\eta)
		}
	}{
		\left[
		1-
		e^{
			-\frac{2\pi W\eta}{\omega_0}
			-\frac{2\pi W_0}{\omega_0}(1-\eta)
		}
		\right]^2
	}.
	&&
\end{flalign}
Here $P_\perp\equiv P_\perp(\eta T)$. 
\begin{figure}[h!]
	\centering
	\includegraphics[width=\linewidth]{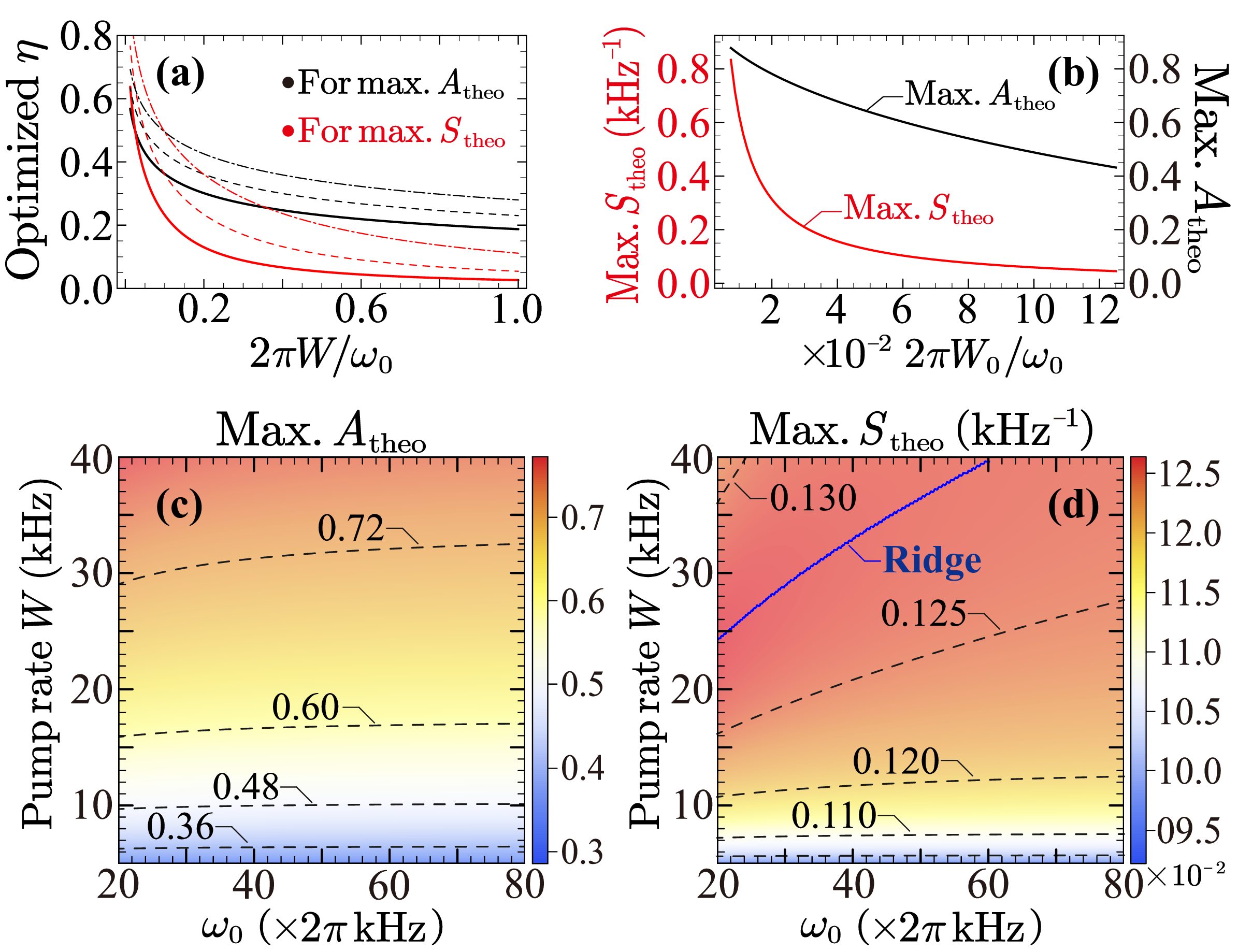}
	\caption{Theoretical duty-cycle optimization of the atomic response. (a) Duty cycles optimizing $A_{\mathrm{theo}}$ (black) and $S_{\mathrm{theo}}$ (red), obtained from the stationary conditions $\partial A_{\mathrm{theo}}/\partial\eta=0$ and $\partial S_{\mathrm{theo}}/\partial\eta=0$ at $\omega_0=2\pi\times40\,\mathrm{kHz}$. Thick solid, thin dashed, and thin dot-dashed curves correspond to $W_0=0.025\omega_0/(2\pi)$, $0.050\omega_0/(2\pi)$, and $0.100\omega_0/(2\pi)$, respectively. (b) Optimized $A_{\mathrm{theo}}$ and $S_{\mathrm{theo}}$ versus $W_0$ at $W=28\,\mathrm{kHz}$. (c) Optimized $A_{\mathrm{theo}}$ over the $(\omega_0/2\pi,W)$ plane at $W_0=2\,\mathrm{kHz}$; dashed curves are contours. (d) Corresponding map of $S_{\mathrm{theo}}$. The blue ridge gives the value of $W$ maximizing $S_{\mathrm{theo}}$ at each $\omega_0$, identifying the optimal pump-on relaxation rate for sensitivity optimization.}
	\label{Fig:theorysummarized}
\end{figure}

We now use Eq.~\ref{eqn:Slopeideal} as a analytical tool to explore the theoretical limit of duty-cycle optimization. In particular, we examine how the $\eta$ and the corresponding maximum response depend on the parameter set \{$W$, $W_0$, $\omega_0$\}. Fig.~\ref{Fig:theorysummarized}(a) compares the duty cycles that maximize $A_{\rm theo}$ and $S_{\rm theo}$ as functions of $2\pi W/\omega_0$. Both optima shift toward smaller $\eta$ as $W$ increases, reflecting that a shorter pump-on duration is sufficient when the optical pumping is stronger. For small $2\pi W/\omega_0$, the two optimal-$\eta$ curves are close and may intersect. As $2\pi W/\omega_0$ increases, however, the two optima separate, with the optimum of $S_{\rm theo}$ shifting toward smaller duty cycles than that of $A_{\rm theo}$. For a fixed $2\pi W/\omega_0$, a smaller $W_0$, corresponding to a longer $T_2$, generally shifts the optimal duty cycle to smaller values. Physically, when the transverse spin polarization survives longer during the pump-off stage, a shorter pump pulse is sufficient to maintain the optimized periodic response. 

Fig.~\ref{Fig:theorysummarized}(b) shows the optimized $A_{\rm theo}$ and $S_{\rm theo}$ as functions of $W_0$ at $W=28\,\mathrm{kHz}$. Both of them decrease as the coherence time $T_2=1/W_0$ shortens, demonstrating that transverse relaxation limits the periodic spin response. The slope decreases more rapidly because it depends on both the response amplitude and the phase dispersion, the latter being particularly sensitive to coherence loss. Accordingly, due to $\delta B_{\rm min}\propto N_V/(\gamma S_{\rm theo})$, the sensitivity deteriorates more rapidly with $W_0$ than the amplitude alone. 

Since the pump-on evolution depends on $W\eta T=2\pi W\eta/\omega_0$, we extend the analysis to the $(\omega_0,W)$ parameter space. Fig.~\ref{Fig:theorysummarized}(c) maps the duty-cycle-optimized on-resonance amplitude $A_{\rm theo}$ at $W_0=2,\mathrm{kHz}$. Over the range considered, the optimized amplitude increases monotonically with $W$, while the dashed contours show that maintaining a given amplitude at higher $\omega_0$ requires a modest increase in $W$. Fig.~\ref{Fig:theorysummarized}(d) provides a more direct guide for sensitivity-oriented optimization by showing the optimized slope $S_{\rm theo}$. Unlike $A_{\rm theo}$, $S_{\rm theo}$ varies nonmonotonically with $W$ at fixed $\omega_0$. Its maximum traces an optimal ridge in the $(\omega_0,W)$ plane, highlighted by the blue curve in Fig.~\ref{Fig:theorysummarized}(d), and the optimal $W$ increases with $\omega_0$. This behavior reflects competition between enhanced spin polarization and increased pump-induced transverse relaxation, the latter suppresses the phase dispersion. Furthermore, the relatively weak variation of $S_{\rm theo}$ around the ridge indicates a broad pump-rate tolerance once $\eta$ is optimized.

\section{\label{eqn:hardwareLIA}Experimental Lock-in Results}
To validate the model under practical demodulation, we used the Moku:Lab lock-in amplifier. Its PLL generated a phase-adjustable sinusoidal local oscillator locked to the pump TTL reference. For each $\eta$, the demodulation phase was adjusted to null the resonant output and then fixed while scanning $\delta\omega_m$ to obtain the dispersive response.

\begin{figure}[h]
	\centering
	\includegraphics[width=\linewidth]{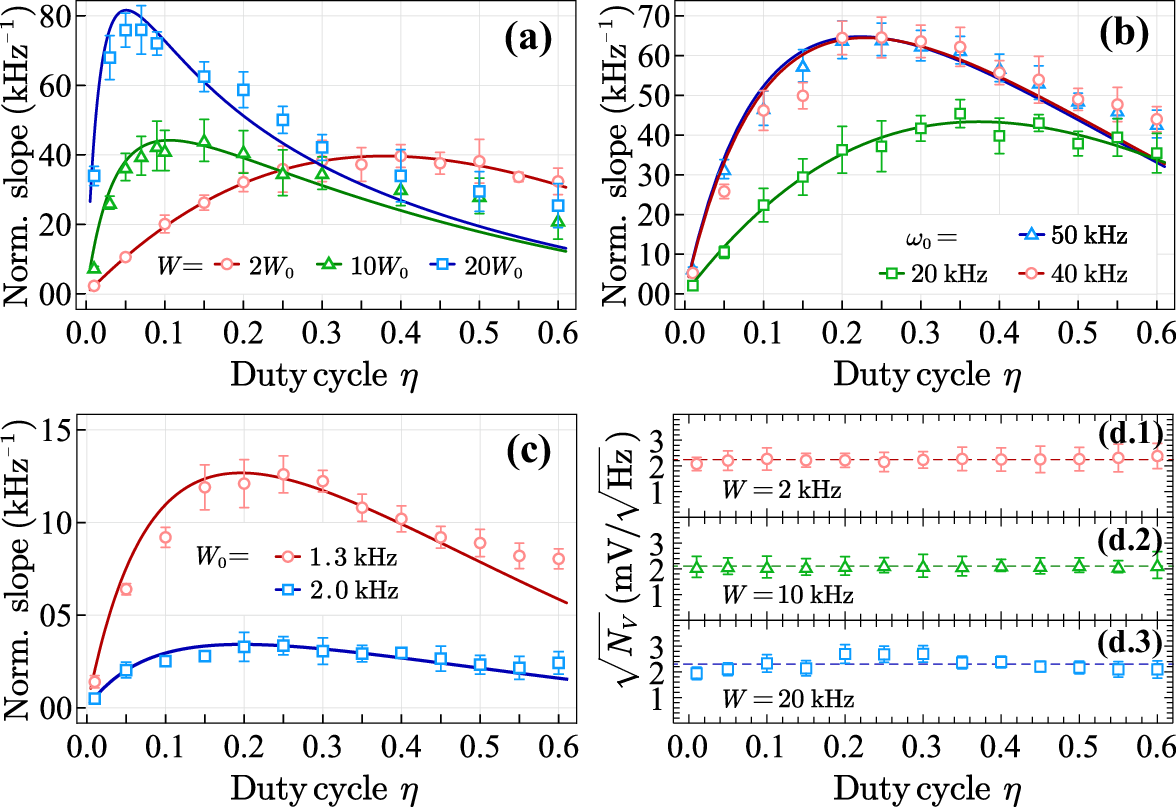}
	\caption{Experimental verification of the duty-cycle dependence of the near-resonance LIA response under different conditions. (a) Results for different total transverse relaxation rates. (b) Results for different bias fields with $W_0=1.1~\mathrm{kHz}$ and $W/\omega_0=0.1/(2\pi)$ kept fixed. (c) Results for different intrinsic transverse relaxation rates with $\omega_0=2\pi\times40~\mathrm{kHz}$ and $W/W_0=5$. The solid curves in (a)–(c) are calculated from the analytical model Eq.~\ref{eqn:Slopeideal}, with a single overall scale factor fitted to each data set. (d.1)–(d.3): Noise amplitude spectral density measured as a function of duty cycle.}
	\label{Fig:MokuData}
\end{figure}

The dependence of the near-resonant LIA slope on the pump duty cycle is summarized in Fig.~\ref{Fig:MokuData}. As shown in Fig.~\ref{Fig:MokuData}(a), the optimal $\eta$ shifts systematically toward smaller values with increasing pump-on relaxation rate $W$. This behavior follows directly from the analytical expression Eq.~\ref{eqn:Slopeideal}: increasing $W$ accelerates spin repolarization during the pump-on interval, so that a shorter pulse is sufficient to establish the transverse polarization, while further increasing the duty cycle primarily enhances pump-induced relaxation. The solid curves, calculated from the analytical model with only an overall scale factor, reproduce both the position of the optimum and its systematic dependence on $W$.

Fig.~\ref{Fig:MokuData}(b) examines the role of the bias field while keeping $W_0=1.1~\mathrm{kHz}$ and $W/\omega_0=0.1/(2\pi)$ fixed. Under this constraint, the pump-induced relaxation accumulated over one Larmor cycle remains approximately unchanged, whereas the relative contribution of the intrinsic relaxation, $W_0/\omega_0$, decreases as the bias field increases. Consequently, the optimum duty cycle shifts to smaller values and the response approaches a common high-field behavior. This illustrates that the duty-cycle optimization is governed not by the absolute relaxation rates alone, but by their ratios to the Larmor frequency, as explicitly encoded in the exponential factors of the analytical expression.

The influence of intrinsic transverse relaxation is further isolated in Fig.~\ref{Fig:MokuData}(c). Here, $\omega_0=2\pi\times40~\mathrm{kHz}$ and $W/W_0=5$. As predicted by Fig.~\ref{Fig:theorysummarized}(b), increasing $W_0$ strongly reduces the maximum slope, reflecting the reduced coherent spin accumulation over successive modulation periods. In contrast, the position of the optimum changes only moderately because the pump-on and pump-off relaxation rates are scaled together. The model agrees well with experiment for $\eta<0.5$, while systematic deviations at larger duty cycles indicate the omitted effect of light-induced relaxation in the present model. Nevertheless, within its range of validity, the agreement across the three parameter scans confirms that the $\eta$-dependence is governed primarily by the competition between polarization buildup during pumping and transverse-coherence preservation during the remaining cycle. 

Finally, Figs.\ref{Fig:MokuData}(d.1)–(d.3) show that the measured noise amplitude spectral density remains nearly independent of duty cycle for the three representative relaxation conditions. Therefore, within the explored parameter range, the duty-cycle dependence of the magnetometric performance is governed mainly by the variation of the response slope rather than by a corresponding change in the detection noise. This establishes the analytical slope maximum as a practical criterion for optimizing the modulation duty cycle.

\section{\label{sec:conclusion}Conclusion}
In conclusion, we have developed a quantitative framework for the duty-cycle-dependent atomic response under full-depth intensity PWM. The model solves the pump-on/off Bloch dynamics subject to a periodic boundary condition, while an extended FID protocol independently calibrates the effective relaxation rates $W$ and $W_0$. Together, they provide a predictive description of the periodically driven spin response.

The model is validated by time-domain waveforms, fast virtual lock-in analysis, and hardware lock-in measurements. Under the present electronic-noise-limited conditions, the near-resonance lock-in slope serves as a practical figure of merit for sensitivity optimization. A compact small-signal expression accurately reproduces the optimal duty cycle and enables systematic optimization over the relaxation rate, pump rate, and Larmor frequency. Notably, each Larmor frequency has an optimal pump rate, demonstrating that stronger pumping does not necessarily improve sensitivity. Future work may incorporate light shifts and spatially inhomogeneous diffusive dynamics \cite{diffusivemodesinthermalgases}, extend the optimization to fundamental-noise-limited operation \cite{gradiometer22,SqueezedLightEnhancement2021}, and apply the framework to miniature vapor-cell platforms \cite{miniature24}.

\section*{Data Availability}
The data that support the findings of this study are available from the corresponding author upon reasonable request.
\begin{acknowledgements}
We acknowledge the fundings from National Natural Science Foundation of China (Grant Nos. 12304561, 12304391, 12447128 and 92576209) and Anhui Province Science and Technology Innovation Overcoming Plan Project (Grant No. 202423s06050002).
\end{acknowledgements}

\bibliography{PWMmagnetometer.bib}

\end{document}